\documentclass[a4paper,fleqn,usenatbib]{mnras}
\usepackage{amssymb,amsmath,graphicx,psfrag,mathtools}
\def\app#1#2{%
	\mathrel{%
		\setbox0=\hbox{$#1\approx$}%
		\setbox2=\hbox{%
			\rlap{\hbox{$#1\propto$}}%
			\lower1.1\ht0\box0%
			}%
			\raise0.25\ht2\box2%
			}%
			}
			\def\approxprop{\mathpalette\app\relax}

\usepackage[T1]{fontenc} 
\usepackage{ae,aecompl} 
\usepackage{url}
\usepackage{revsymb}
\hypersetup{draft} 
\title[Solar Radio-Frequency Reflectivity]{Solar Radio-Frequency
Reflectivity and Localization of FRB from Solar Reflection
}
\author[S. Wang \& J. I. Katz]{
	S. Wang$^{1}$ \& J. I. Katz,$^{2}$\thanks{E-mail katz@wuphys.wustl.edu} 
\\
$^{1}$Department of Physics, Washington University, St. Louis, Mo. 63130
USA
\\
$^{2}$Department of Physics and McDonnell Center for the Space Sciences,
Washington University, St. Louis, Mo. 63130 USA 
}
\date{Accepted XXX.  Received YYY; in original form ZZZ} 
\pubyear{2022} 
\date{\today}
\begin{document} 
\label{firstpage} 
\pagerange{\pageref{firstpage}--\pageref{lastpage}} 
\maketitle 
\begin{abstract}
	The radiation of a Fast Radio Burst (FRB) reflects from the Moon
	and Sun.  If a reflection is detected, the time interval between the
	direct and reflected signals constrains the source to a narrow arc on
	the sky.  If both Lunar and Solar reflections are detected these two
	arcs intersect, narrowly confining the location on the sky.  A
	previous paper calculated reflection by the Moon.  Here we calculate
	the reflectivity of the Sun in the ``flat Sun'' approximation as a
	function of angle of incidence and frequency.  The reflectivity
	is high at low frequencies ($\lessapprox 100\,$MHz) and grazing
	incidence (angles $\gtrapprox 60^\circ$), but exceeds 0.1 for
	frequencies $\lessapprox 80\,$MHz at all angles.  {However,
	the intense thermal emission of the Solar corona likely precludes
	detection of the Solar reflection of even MJy Galactic bursts like
	FRB 200428.}
\end{abstract}
\begin{keywords} 
radio continuum, transients: fast radio bursts, Sun: atmosphere
\end{keywords} 
\section{Introduction}
The all-sky FRB rate, above a threshold $\sim 1$ Jy-ms at 1400 MHz, is $\sim
10^6$/sky-year \citep{CC19,PHL22}.  Despite this, until recently only $\sim
100$ distinct FRB sources had been observed \citep{P16,TNS} because most
radio telescopes have very limited fields of view.  For example, an
individual Parkes beam has a width of about $10^{-5}$ sterad at this
frequency; its 13 beams together cover about $10^{-5}$ of the sky.
CHIME/FRB \citep{CHIME,C21a} has the comparatively large field of view of
200 square degrees and discovered 536 distinct FRB sources above a 400--800
MHz fluence threshold of about 5 Jy-ms in about a year \citep{C19b,F20,C21b}.
STARE2, consisting of a network (providing interferometric localization
information) of choke-ring (essentially dipole) feeds \citep{B20a}, has a
field of view of about 3.6 sterad, about 30\% of the sky, at the price of 
the very high L-band detection threshold of $\sim 300$ kJy-ms.

It has long been realized \citep{K14} that a ``cosmological'' FRB in our
Galaxy would be bright enough to be observed by a single half-wave dipole
antenna, and that a small network of such dipoles could localize it.  STARE2
observed \citep{B20b} the first (at the time of writing, only) Galactic FRB
200428, even though it was less energetic than any observed extra-Galactic
FRB.  It was also, more fortuitously, observed by CHIME/FRB \citep{C20a}.
More FRB with accurate positions, as well as observations with full-sky
sensitivity, could identify future Galactic FRB.

A radio telescope whose beam is {approximately} matched to the angular
size of the Moon or the {somewhat larger \citep{MC15} radio size of the}
Sun (telescope diameter about 22 m in L-band) and staring at that object
could detect reflected radiation from a FRB anywhere in the sky, with
sensitivity about one order of magnitude less than that of a dipole or a
single element of STARE2.  Although insufficiently sensitive to detect FRB
at cosmological distances, this was proposed \citep{K20} as a method to
detect Galactic micro-FRB, and anticipated the the first such event
discovered, FRB 200428.

Greater sensitivity could be provided by a larger telescope with a
multi-beam feed covering the Moon or Sun.  Comparing the phases of the
signals from such telescopes with that from STARE2 or a similar instrument
would provide two very long interferometric baselines, one equal to the
projected Earth-Moon separation and one equal to the projected Earth-Sun
separation, and would therefore enable precise localization on the sky.  At
the lower frequencies at which the Solar reflectivity is high, the
resolution-matched telescope would be much larger than 22 m and its
sensitivity higher, but even a 22 m telescope could have useful sensitivity.

\begin{figure}
	\centering
	\includegraphics[width=3in]{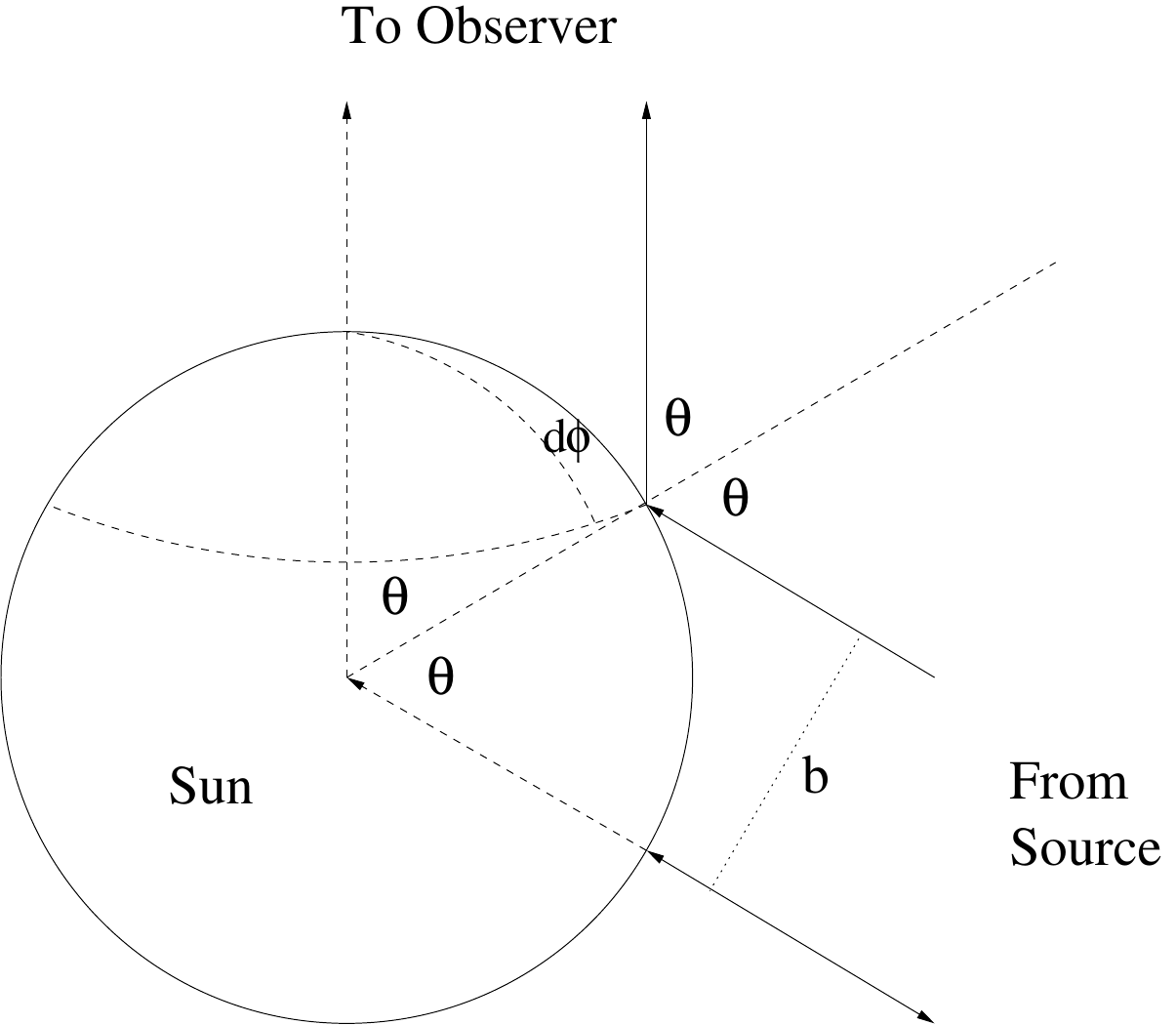}
	\caption{\label{sunscatt} Scattering of FRB radiation by the Sun
	(after Fig.~1 of \citet{K20}; {asymptotes of the ray paths are
	shown}).  The angle of incidence is $\theta$.}
\end{figure}

This paper calculates the reflectivity of the Sun as a function of frequency
and angle of incidence using a known model of the Solar atmosphere and
corona.  The reflectivity must be known to evaluate the feasiblity of 
observing FRB reflected by the Sun.  The geometry is shown in
Fig.~\ref{sunscatt}.  {Qualitative refracted ray paths were shown by
\citet{N61} before the development of modern quantitative coronal models.}
\section{Reflectivity}
\label{reflectivity}
\citet{K20} estimated that the radio-frequency reflectivity of the Sun is
low at frequencies $\gtrsim 200\,$MHz, but high at lower frequencies.  FRB
have not generally been observed at low frequencies, in part because
dispersion delays and scattering broadening are much greater at low 
frequencies, the former scaling $t_{dispersion}\propto \nu^{-2}$ (and the
derivative that measures the differential arrival time across a channel
$dt_{dispersion}/d\nu \propto \nu^{-3}$), and $dt_{scatter}/d\nu
\approxprop \nu^{-4}$.  However, FRB 20180916B was detected by LOFAR at
frequencies as low as 120 MHz \citep{PM20,P21}, demonstrating that at least
some FRB may be observed at these low frequencies at which the Solar
reflectivity is expected to be high.

Radar measurements of the Solar reflectivity only measure it at normal
incidence, but most FRB specularly reflected by the Sun to the Earth will
have angles of incidence far from normal, so their rays do not penetrate the
denser and more absorptive lower layers of the Solar atmosphere.

Radio telescopes operating at lower frequencies ($< 300$ MHz), such as the
Murchison Widefield Array \citep{T13,Wa18}, LOFAR \citep{vH13} and the
planned SKA LFAA \citep{dLA20}, are generally phased arrays of dipoles.
Focusing is electronic (beams are synthesized) so the telescope can
effectively ``stare'' at the Sun without imposing a focussed visible and
infrared light heat load on the receiving elements.  The older Giant
Metrewave Radio Telescope \citep{A95} uses parabolic dishes to focus radio
waves, but its dishes are made of an open (7\% filled) wire mesh that does
not efficiently focus visible light because the wires are cylindrical and
much thicker than the wavelength of visible light.   If necessary, the
receiver can be protected from focussed sunlight with a thin sheet of opaque
(to visible and infrared light) plastic.

Synthesized beams have complex angular structure, rather than being simply
matched to the angular size of the Sun, as would be possible for a parabolic
dish at shorter wavelengths and proposed for parabolic reflectors staring
at the Moon at higher frequencies.  Despite this, they are sensitive to FRB
scattered by the Sun; aside from radio frequency interference (generally
narrowly confined in frequency), there is little radiation with the temporal
characteristics of FRB in any direction, other than FRB themselves.  At low
frequencies the dispersion and scatter broadening characteristic of FRB are
large, making them easy to discriminate from any other transients in a
synthesized beam.  Solar-scattered FRB would be distinguishable from
unscattered FRB simultaneously observed in other lobes of the beam by
different dispersion measures and by the possible earlier direct path
observation of flux from the FRB by instruments like STARE2.

The reflectivity
\begin{equation}
	\label{R}
	{\cal R}(\nu,\theta) = \exp{(-\tau(\nu,\theta))},
\end{equation}
where $\nu$ is the wave frequency and $\theta$ is determined by the
direction to the FRB.  The absorption optical depth along the ray path
\begin{equation}
	\label{tau}
	\tau(\nu,\theta) = \int\!\kappa(\nu,n_e({\vec r}),T({\vec r}))\,ds,
\end{equation}
where $\kappa(\nu,n_e,T)$ is the opacity, the electron density $n_e =
n_e({\vec r})$ and the temperature $T({\vec r})$ are found from a model
{\citep{N67}} of the Solar corona.

The path ${\vec r}(s)$ of a ray of radiation is found, in the geometrical
optics limit, from the eikonal equation \citep{B04}
\begin{equation}
	\label{eikonal}
	{d\phantom{0} \over ds} \left({n{d{\vec r} \over ds}}\right) = 
	{\vec \nabla} n,
\end{equation}
where $ds$ is an element of path length, the refractive index
\begin{equation}
	n = \sqrt{1 - {\omega_p^2 \over \omega^2}},
\end{equation}
$\omega = 2 \pi \nu$ and the plasma frequency
\begin{equation}
	\omega_p = \sqrt{4 \pi n_e e^2 \over m_e}.
\end{equation}

For a specified $n_e({\vec r})$, we integrate Eq.~\ref{eikonal} numerically
to find the path.  {In a ``flat Sun'' approximation the initial
condition is the angle of incidence $\theta$ at an altitude far above the
Solar surface, where $\omega_p \ll \omega$ and refraction is negligible.
This is related to the impact parameter $b$ by} $b = b(\theta)$.  In general
this relation is non-trivial, but in a ``flat Sun'' approximation $b = R
\sin{\theta}$, where $R$ is the Solar radius and $\vec r$ is replaced by an
altitude $z$.  This approximation is justified because a 150 MHz ray
at near-normal incidence penetrates, {fitting the profile of
\citet{N67}}, to an altitude {$\approx 0.04 R_\odot$} above the
photosphere, where the scale height of the Solar atmosphere {$\approx
0.09 R_\odot$}.  At shallower angles of incidence the penetration is
shallower, the attenuation less, and its quantitative calculation less
important.

{There is an extensive literature on the properties of the Solar corona
({\it e.g.\/} \citet{S77,G14,M18,Z21}).  Unfortunately, only a fraction of
this literature is concerned with the {background Solar corona} (as
opposed to phenomena such as coronal streamers {or Type III radio
bursts}) and apparently none provides both the density and
the temperature in the inner coronal region relevant here.  Instead, we fit
the smooth density profile of an isothermal quiet Solar corona
\begin{equation}
	n_e(r) = n_0 \exp{\left({GM\mu \over k_B T r}\right)}
\end{equation}
to the classic results of \citet{N67}, where the molecular weight $\mu$ per
particle is taken as $0.6 m_p$, where $m_p$ is the proton mass, appropriate
to fully ionized hydrogen and helium.  A good fit is obtained for $T = 1.1
\times 10^6\,$K, a value we adopt to calculate the opacity. 

More detailed modern studies \citep{MC15} show that even the quiet corona is
both very variable and not spherically symmetric.  Because the present work
is only a feasibility study whose purpose is to estimate whether
Solar-reflected FRB may be observable, we do not need very accurate
temperature and density profiles; the variability of the corona makes any
model profile only approximately applicable at any particular time and
for any particular direction to a FRB.  If reflected FRB are observed the
quantity of interest will be the time delay between the direct and reflected
signals, not their comparative amplitudes.}

The opacity \citep{S62}, including the effect of stimulated emission,
\begin{equation}
	\label{kappa}
	\begin{split}
		\kappa(\nu,n_e,T) = &{4 \over 3}\sqrt{2 \pi \over 3 k_B T}
	{n_e \sum_Z n_Z Z^2 e^6 \over h c m_e^{3/2} \nu^3}\\
		&\left\{1-\exp{\left[-\left({h\nu \over k_B T}\right)\right]}
		\right\}g_{ff}\\ &\approx {4 \over 3}\sqrt{2 \pi \over 3}
		{n_e \sum_Z n_Z Z^2 e^6 \over (k_B T m_e)^{3/2} c \nu^2}
		g_{ff},
	\end{split}
\end{equation}
where $n_Z$ is the density of ions with charge $Z$ (allowing for multiply
ionized atoms in the hotter regions), $k_B$ is the Boltzmann constant and
$g_{ff}$ is the Gaunt factor.  The Gaunt factor depends only logarithmically
on the parameters, except where $\omega \to \omega_p$ and the group velocity
is significantly less than $c$ \citep{S62}, and is typically about 15
\citep{vH14}.  Rays whose angles of incidence are not small do not closely
approach $\omega = \omega_p$.  Eqs.~\ref{R}, \ref{tau} are then used to find
the reflectivity.

\begin{figure}
	\centering
	\includegraphics[width=3.3in]{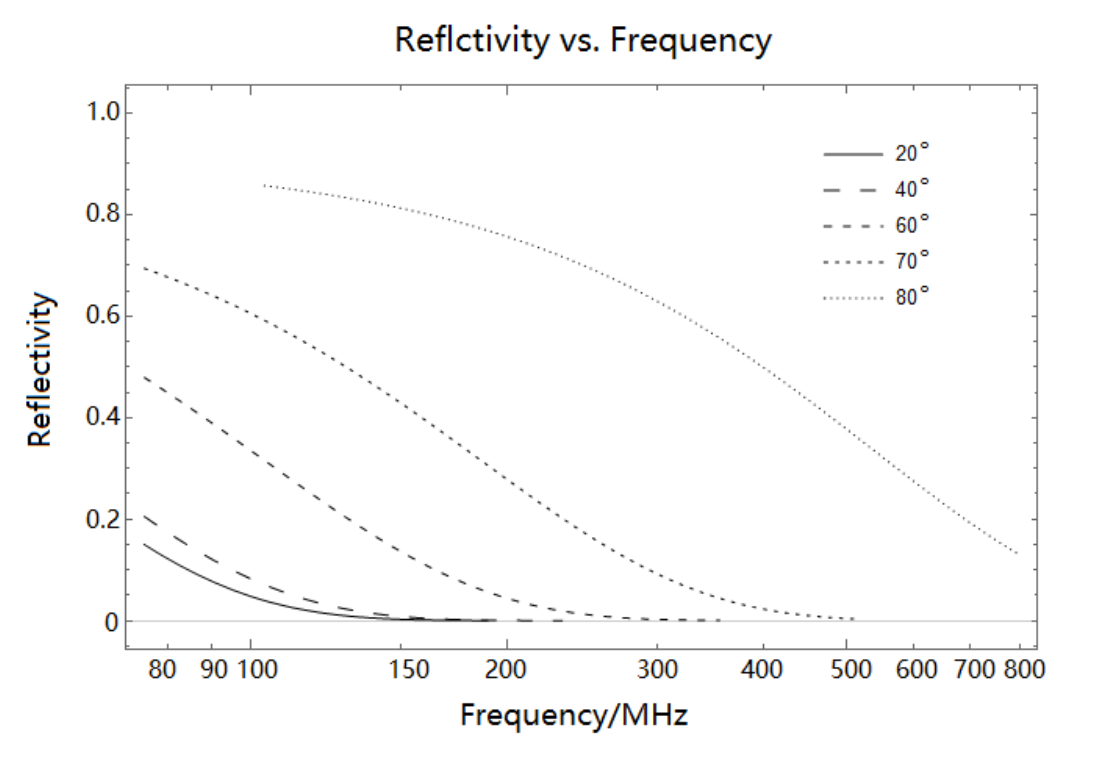}
	\caption{\label{freq} Solar reflectivity {\it vs.\/} frequency
	at several angles of incidence $\theta$.}
\end{figure}

\begin{figure}
	\centering
	\includegraphics[width=3.3in]{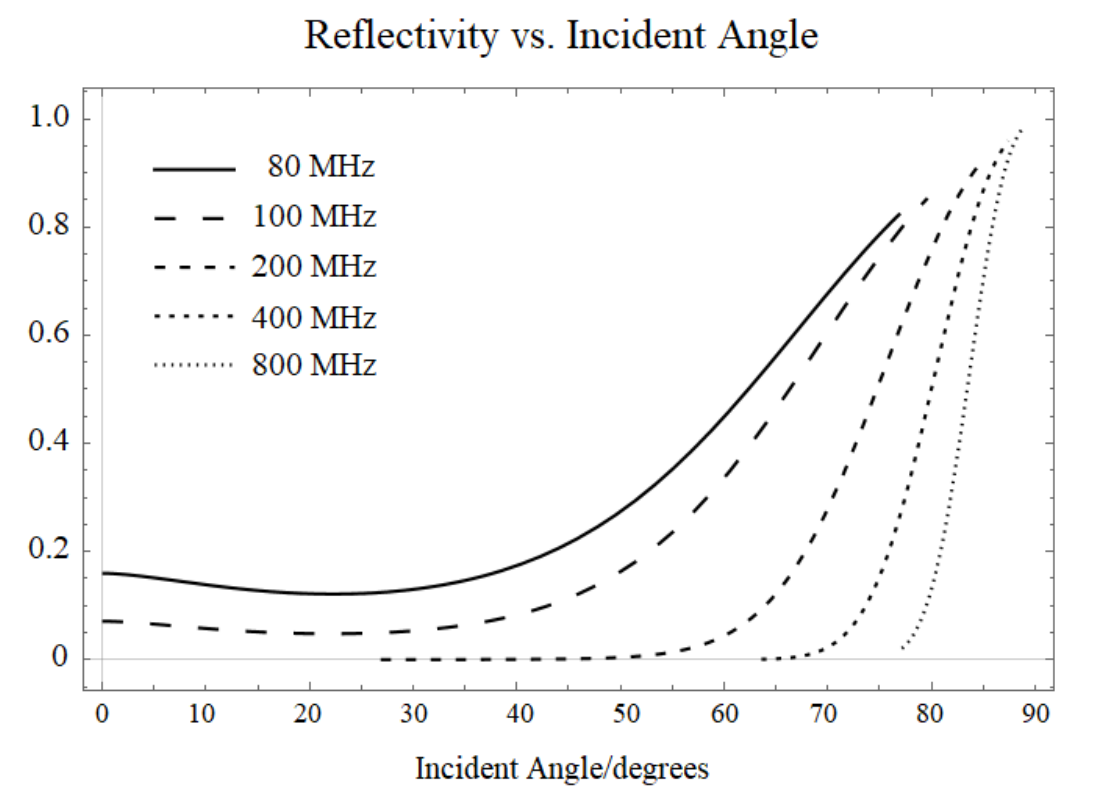}
	\caption{\label{angle} Solar reflectivity {\it vs.\/} angle of
	incidence at several frequencies.}
\end{figure}

The results for several angles of incidence are shown as functions of the
frequency in Fig.~\ref{freq} and as functions of the angle of incidence
in Fig.~\ref{angle}.
\section{Solar Background}
In addition to scattered FRB, a telescope pointed at the Sun receives
Solar radiation, with which the FRB signal must compete.  At the frequencies
of interest, 100--200 MHz, the {quiet Sun is nearly a thermal emitter at
its coronal temperature with a brightness temperature} $T_b \sim 10^6\,$K,
corresponding to a flux $F_{b,\nu} \sim 10^4\,$Jy \citep{K66}.  During
periods of Solar activity $T_b$ may be much higher; scattered FRB cannot be
detected at such times.  If the Sun fills the field of view (or the
resolution element of a larger telescope), the background quiet Sun antenna
temperature $T_b \sim 10^6\,$K.  A telescope staring at the Moon has $T_b
\approx 235\,$K, its mean surface temperature.

A FRB (or other distant source) with direct flux $F_\nu$ produces a scattered
flux at the Earth \citep{K20}
\begin{equation}
	\label{Fscatt}
	F_{scatt} = {{\cal R}(\theta,\nu) \over 4}{R^2 \over D^2}
	\cos{\theta} F_\nu \approx 5 \times 10^{-6} {\cal R}(\theta,\nu)
	\cos{\theta} F_\nu,
\end{equation}
where ${\cal R}(\theta,\nu)$ is the reflectivity calculated in
Sec.~\ref{reflectivity}, $\theta$ is the angle of incidence, $R = R_\odot$
the radius of the scattering object, $D = 1\,$AU its distance {and
$F_\nu$ is the unscattered flux ($R/D$ is nearly the same for the Moon and
the Sun).

For a telescope matched to the angular size of the Sun with diameter $d =
\lambda D/2R$, the antenna temperature produced by the scattered radiation
\begin{equation}
	\begin{split}
		k_B T_{scatt} &= {\pi \over 4} d^2 F_{scatt}\\ &=
		{\pi \over 4} d^2 {{\cal R}\cos{\theta} \over 4}
		\left({R \over D}\right)^2 F_\nu\\ &= 
	{\pi \over 16} {\cal R} \cos{\theta} \lambda^2 F_\nu
	\end{split}
\end{equation}
or
\begin{equation}
	T_{scatt} \approx 1300\ {\cal R} \cos{\theta} \left({\text{100 MHz}
	\over \nu}\right)^2 {F_\nu \over \text{1 MJy}}\ \text{K}.
\end{equation}
For a matched telescope the antenna temperature $T_{scatt}$ is independent
of the angular size of the scatterer.  At $\nu = 100\,$MHz such a telescope
would have $d \approx 350\,$m, which is not feasible.  For a more feasible
telescope with $d = 25\,$m
\begin{equation}
	T_{scatt} \approx 7\ {\cal R} \cos{\theta} {F_\nu \over
	\text{1 MJy}}\ \text{K}.
\end{equation}

The limiting factor in observations in the direction of the Sun is not the
receiver noise temperature but the thermal emission from the corona, that
may be approximated as a black body of temperature $T_{corona} \sim 10^6
(1-{\cal R})\,$K, where the physical temperature $\sim 10^6\,$K
is multiplied by the emissivity $1 - {\cal R}$.  Integrated over a bandwidth
$B$ and a time (such as the duration of a FRB) $\Delta t$, the signal to
(background) noise ratio of detection
\begin{equation}
	\label{S/N}
	\begin{split}
		{S \over N} &\approx {T_{scatt} \over T_{corona}}
		\sqrt{B \Delta t}\\ &\sim 0.003 {F_\nu \over 1\,\text{MJy}}
		\sqrt{B \Delta t \over 100\,\text{MHz-ms}}
		{{\cal R} \over 1-{\cal R}}.
	\end{split}
\end{equation}
In addition, scattering in the turbulent corona would temporally broaden a
FRB, reducing $F_\nu$ (more than it would increase $\Delta t$), $T_{scatt}$
and $S/N$.  \citet{Z21} estimate scattering widths $\Delta t \sim 0.5\,$s at
150 MHz, that would reduce $S/N$ by an additional factor $\sim 20$ in
comparison to an unbroadened $\Delta t \sim 1\,$ms.

It is evident that a Galactic FRB would have to have flux $\gtrsim
\text{GJy}$ for its Solar reflection to be detectable above the coronal
thermal emission.  Although such an event would be unprecedented, a 1 Jy FRB
at $z \sim 1$ would have a flux $\sim 100 GJy$ at 10 kpc, indicating that
such events may occur.  Events producing 0.01--0.1 Jy at ``cosmological''
distances would be 1--10 GJy at Galactic distances, are more frequent than
the most extreme events, and their Solar reflections are likely detectable.}
\section{Discussion}
Figs.~\ref{freq} and \ref{angle} show that for frequencies $\nu \gtrapprox
150\,$MHz the reflectivity the reflectivity is substantial only for angles
of incidence $\theta \gtrapprox 60^\circ$.  The fraction of the sky with
angles of incidence $\ge \theta$ is $(1+\cos{2\theta})/2$, which is 0.25
for $\theta = 60^\circ$ but only 0.12 for $\theta = 70^\circ$.  Unlike the
Moon, the Sun is a far-from-isotropic reflector at those frequencies and
effectively reflects only a fraction of isotropically distributed sources,
such as FRB.  At lower frequencies $\nu \lessapprox 100\,$MHz, the Sun is a
good reflector, better than the Moon \citep{K20}, at all angles of
incidence.  

{The results of this study were disappointing because of the intense
background of Solar coronal thermal emission.  This is in contrast to the
results of our earlier study of Lunar scattering; the Lunar surface is only
a weak source of thermal emission at radio frequencies.}
\section*{Acknowledgement}
{We thank an anonymous reviewer for several rounds of constructive
criticism and for pointers to the Solar physics literature with which we
were unfamiliar.  {In particular, we appreciate his calling to our
attention the intense thermal coronal background that determines the
observability of Solar reflections.}}
\section*{Data Availability}
This theoretical study did not generate any new data.

\label{lastpage}
\end{document}